\documentclass{article}

\usepackage{nips_2017}

\usepackage[utf8]{inputenc} 
\usepackage[T1]{fontenc}    
\usepackage{hyperref}       
\usepackage{url}            
\usepackage{booktabs}       
\usepackage{amsfonts}       
\usepackage{nicefrac}       
\usepackage{microtype}      

\usepackage{comment}
\usepackage{graphicx}  
\usepackage{subcaption}
\usepackage{algorithm}
\usepackage[noend]{algpseudocode}
\usepackage[usenames, dvipsnames]{color}

\title{Newtonian Action Advice: Integrating Human Verbal Instruction with Reinforcement Learning}

\author{
  Samantha Krening\\
  Institute for Robotics and Intelligent Machines\\
  Georgia Institute of Technology\\
  Atlanta, GA 30332 \\
  \texttt{skrening@gatech.edu} \\
   \And
   Karen M. Feigh \\
   Institute for Robotics and Intelligent Machines \\
   Georgia Institute of Technology \\
   \texttt{karen.feigh@gatech.edu} \\
}

\begin{document}

\maketitle

\begin{abstract}

A goal of Interactive Machine Learning (IML) is to enable people without specialized training to teach agents how to perform tasks. Many of the existing machine learning algorithms that learn from human instructions are evaluated using simulated feedback and focus on how quickly the agent learns. While this is valuable information, it ignores important aspects of the human-agent interaction such as frustration. In this paper, we present the Newtonian Action Advice agent, a new method of incorporating human verbal action advice with Reinforcement Learning (RL) in a way that improves the human-agent interaction. In addition to simulations, we validated the Newtonian Action Advice algorithm by conducting a human-subject experiment. The results show that Newtonian Action Advice can perform better than Policy Shaping, a state-of-the-art IML algorithm, both in terms of RL metrics like cumulative reward and human factors metrics like frustration. 

\end{abstract}





A goal of interactive machine learning is to enable people who are not ML experts to naturally and intuitively teach agents how to perform tasks. This paper introduces an algorithm, Newtonian Action Advice, which incorporates a human's verbal action advice with Reinforcement Learning (RL). The algorithm leverages a simple physics model to provide an agent that acts in a way people expect and find non-frustrating.

We validate our algorithm by first constructing oracles to simulate human feedback to compare Newtonian Action Advice (NAA) with Policy Shaping and Bayesian Q-learning. In addition to the simulations, we conducted a human-subject experiment in which participants trained both NAA and Policy Shaping agents, and then reported on the experience of working with both agents. 


We suggest that validating interaction algorithms with oracles and analyzing traditional RL metrics such as cumulative reward and training time is only the first step. In addition to RL metrics, interaction algorithms should be validated by measuring peoples' experiences with agents using human factors, such as frustration. Ideally, the interaction algorithms should be designed with the goal of creating a positive human experience, because individuals who experience frustration while interacting with an agent are unlikely to continue or repeat the interaction in the future.



The results show that Newtonian Action Advice can perform better than Policy Shaping, both in terms of RL metrics like cumulative reward and human factors metrics like frustration. NAA can learn faster using less human instruction than Policy Shaping. NAA creates a better human experience than Policy Shaping. Compared to Policy Shaping, participants found the Newtonian Action Advice agent to be less frustrating, clearer and more immediate in terms of how the agent used human input, better able to complete the task as the participant intended, and more intelligent.




\section{Background}

\subsection{Reinforcement Learning}

Reinforcement learning (RL) is a form of machine learning influenced by behavioral psychology in which an agent learns what actions to take by receiving rewards or punishments from its environment~\citep{sutton1998reinforcement,skinner1938behavior}. The probability people will repeat an action in a given circumstance is increased or decreased if they receive positive or negative reinforcement. 

Most RL algorithms are modeled as Markov Decision Processes (MDPs), which learn policies by mapping states to actions such that the agent's expected reward is maximized. An MDP is a tuple $(S,A,T,R,\gamma)$ that describes $S$, the states of the domain; $A$, the actions the agent can take; $T$, the transition dynamics describing the probability that a new state will be reached given the current state and action; $R$, the reward earned by the agent; and $\gamma$, a discount factor in which $0 \le \gamma \le 1$.

Bayesian Q-Learning is an MDP-based RL algorithm in which the utility of state-action pairs are represented as probabilistic point estimates of the expected long term discounted reward \citep{dearden1998bayesian}. Bayesian Q-Learning was used as the underlying RL algorithm for both the Policy Shaping and NAA interaction methods in this work.

\subsection{Learning from critique: Policy Shaping}

Critique was initially used directly as a reward signal~\citep{isbell2001social}, but it was later shown ~\citep{knox2010combining,thomaz2008teachable} that it is more efficient to use critique as policy information. Policy Shaping is an interaction algorithm that enables a human teacher's critique to be incorporated into a Bayesian Q-learning agent as policy information \citep{griffith2013policy} and was used in this work. \citet{cederborg2015policy} investigated how to interpret silence while learning from critique with policy shaping.

\subsection{Learning from advice}

Various forms of advice have been developed in other work, including linking one condition to each action~\citep{maclin2005giving}, and linking a condition to rewards~\citep{MacGlashan-RSS-15}. Several connect conditions to higher-level actions that are defined by the researcher instead of primitive actions~\citep{maclin2005giving, kuhlmann2004guiding, joshi2012object}.~\citet{argall2008learning} creates policies using demonstrations and advice.~\citet{mericcli2014interactive} parses language into a graphical representation and finally to primitive actions.~\citet{maclin2005giving} has the person provide a relative preference of actions.~\citet{sivamurugan2012instructing} explored learning multiple interpretations of instructions. ~\citet{tellex2011understanding} represents natural language commands as probabilistic graphical models.

Most methods are permanently influenced by the advice.~\citet{kuhlmann2004guiding} can adjust for bad advice by learning biased function approximation values that negate the advice.~\citet{maclin2005giving} uses a penalty for not following the advice that decreases with experience. Newtonian Action Advice differs because the advice can be overwritten by new, contradictory advice in the future. 

Many researchers incorporate advice using IF-THEN rules and formal command languages~\citep{maclin2005giving, kuhlmann2004guiding}; if the state meets a condition, then the learner takes the advice into account. Formal command languages and IF-THEN rules require advice that is state specific and contains numbers. Similar to this work, the advice in~\citet{argall2008learning} does not require people to give specific numbers for continuous state variables, but uses a set of predefined advice operators. 


\subsection{Natural Language Processing (NLP)}

\subsubsection{Automatic Speech Recognition (ASR)}
ASR software transcribes the human teacher's verbal instructions to written text. The human-subject experiment in this work used the Sphinx ASR software ~\citep{huggins2006pocketsphinx}.


\subsubsection{Sentiment Analysis}

Sentiment analysis is an NLP tool used to classify movie, book, and product reviews into positive and negative ~\citep{pang2008opinion}. Sentiment analysis has not been widely applied to action selection. One method we previously developed for using sentiment analysis is to classify natural language advice into advice of `what to do' and warnings of `what not to do' ~\citep{krening2017learning}. Many approaches to learning from language instruction require people to provide instructions using specific words, often in a specific order or format~\citep{mericcli2014interactive}. ~\citet{thomason2015learning} worked to get around limitations like keyword search by creating an agent that learns semantic meaning from the human. In this work we created a method of using sentiment analysis to filter verbal critique into positive and negative, which furthers the goal of allowing people to provide verbal instructions without being limited to a specific dictionary of words. 

%

This work uses Stanford's deep learning sentiment analysis software ~\citep{manning-EtAl:2014:P14-5}, which uses Recursive Neural Tensor Networks and the Stanford Sentiment Treebank ~\citep{socher2013recursive}. The Stanford Sentiment Treebank is a set of labeled data corpus of fully-labeled parse trees trained on the dataset of movie reviews from rottentomatoes.com ~\citep{pang2005seeing}.

\section{Newtonian Action Advice}


Newtonian Action Advice is a teaching interaction algorithm we designed to enable an RL agent to learn from human action advice. The theory is a metaphor of Newtonian dynamics: objects in motion stay in motion unless acted on by an external force. In the Newtonian Action Advice model, a piece of action advice provided by the human is an external force on the agent. Once a person provides action advice (ex: ``Go right''), the agent will immediately move in the direction of the external force, superseding the RL agent's normal action selection choice of exploration vs. exploitation. The model contains natural friction that `slows down' the agent's need to follow the human's advice. The friction ensures that after some amount of time, the agent will resume the underlying RL algorithm's exploration policy. The advice does not necessarily need to specify directional motion; the metaphor of advice as a force pushes the agent to follow the advice as opposed to the RL's action selection mechanism. 

\begin{figure}[b]
\centering
\includegraphics[width=0.45\textwidth]{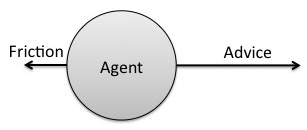}
\caption{Simple Force Model. Actions are an external force acting on the agent, and `friction' determines the amount of time the action will be followed after the advice is given. }
\end{figure}

Newtonian Action Advice was designed to behave in a manner that is more natural and intuitive for the human teacher than other IML algorithms such as Policy Shaping. The force model allows each piece of advice to be generalized through time. If a person says, ``go right,'' the Newtonian Action Advice agent will move right and keep moving right until the `friction' causes the agent to resume normal exploration. The simplicity of the force model is a feature to improve the human experience; people deal with Newtonian mechanics in their everyday life and are used to objects moving in a Newtonian manner. For example, if a ball is thrown straight up in the air the same way multiple times, it will always rise to a certain height and fall back to the ground. The motion is predictable, and one does not need formal physics training to recognize and expect the motion. We expect that loosely mimicking this will help create a user-friendly experience. 

If advice was followed in one state, the Newtonian Action Advice algorithm will cause the agent to follow the same advice if the state is seen in the future. This means that a person will only have to provide advice once for a given situation. Also, only the latest advice is saved for a state, so people can correct any mistakes they made or change the desired policy in real time.

Bayesian Q-Learning was used as NAA's underlying RL algorithm. This choice was primarily made so the Newtonian Action Advice algorithm could be more directly compared to Policy Shaping. The structure of the NAA algorithm is such that the BQ-L algorithm could be exchanged for a different RL algorithm. 





\begin{algorithm}
\caption{Newtonian Action Advice algorithm}\label{alg:newtonianAA}
\begin{algorithmic}[1]
\Procedure{NAA}{}
\For{ each time step}
\State Listen for human advice
\If{human advice given}
\State $newAdvice(state,advice)$
\EndIf
\State $action = actionSelection()$  
\State Take action and get reward
\State Update Bayesian Q-learning policy with reward
\EndFor
\EndProcedure
\Procedure{newAdvice}{$state,advice$}
\State $agent.adviceJustGiven\gets True$
\State $agent.advisedAction\gets advice$
\State $agent.adviceDictionary[state] = advice$
\EndProcedure
\Procedure{actionSelection}{$state$}
\If{$agent.adviceJustGiven == True$}
	\State $chosenAction\gets agent.advisedAction$
    \If{$state \notin agent.adviceDictionary$}
    \State $agent.adviceDictionary[state] = chosenAction$
    \EndIf
    timesNewAdviceFollowed += 1
    \If{$timesNewAdviceFollowed >= timesToFollowAdviceThreshold$}
    \State $agent.adviceJustGiven = False$
    \State $timesNewAdviceFollowed = 0$
    \EndIf
\EndIf
\State If advice has been given for this state, choose between the human advice and the algorithm
\State Otherwise, if advice has not been given for this state in the past, use the action recommended from the BQL algorithm
\State \textbf{return} $chosenAction$

\EndProcedure
\end{algorithmic}
\end{algorithm}


In algorithm \ref{alg:newtonianAA}, \textit{NAA} is the main procedure. At each time step, the agent listens for advice. If advice is given, the agent updates its internal advice dictionary. The agent then chooses and takes an action, receives a reward, and updates the Bayesian Q-learning policy. 

The \textit{New Advice} procedure adds the new (state, advice) pair to the agent's dictionary and sets a parameter that will tell the action selection procedure to follow the new advice. 

The \textit{Action Selection} procedure first checks to see whether advice has recently been given and should still be followed. If the advice is being generalized through time (due to low friction) and the new (state, advice) pair has not been added to the dictionary, it will be added at this time. If this state is revisited in the future, the recent advice given for a previous state will be applied as if it had been given for this state, too. The timer that keeps track of the friction parameter threshold is updated. If the timer indicates that the advice has been followed for long enough, parameters will be reset so the agent will return to the Bayesian Q-learning's action selection for the next time step. If advice has previously been given for this state, the agent must choose between the human advice and the BQL suggestion. For this work, we always choose the human's advice. If a researcher wants to encourage more exploration, a different method can be chosen (i.e. an algorithm similar to $\epsilon-greedy$ applied to human vs. agent action selection instead of exploration vs. exploitation). However, we have found that following advice in a probabilistic manner increases frustration and uncertainty since the agent seems to disregard advice. In the case that no advice has been given for or generalized to the current state, the action is chosen from the Bayesian Q-learning's action selection method.



\subsection{Combining Supervised and Reinforcement Learning to allow for personalization by end-users} \label{subsec:customize}

When designing an interactive machine learning algorithm, one first must ask: what is the goal of IML? Is the goal to use human instruction to decrease training time? Or is the goal to enable people to teach an agent to perform a task in the way the human intends?

If the goal of IML is to use human instruction to decrease the amount of time it takes to train the RL agent, at first glance it seems like the best of both worlds. We get to use powerful RL algorithms that are capable of learning from their environment instead of having policies hard-coded or models from datasets built a priori; and we get to decrease training time by getting useful human input.   

However, if the goal is to get the agent to perform a task that a non-expert specifies in the way the human wants the task done, then we have a problem. 

The policy an RL agent learns is very sensitive to the reward function. In this work, as well as most IML research, the reward function is provided by the researcher prior to the experiment. Given a reward function, an RL algorithm will learn a policy to maximize the reward, but the policy learned may be very alien to a human mind. The agent will technically complete the task efficiently, but not in a way that makes a lot of sense to a human. Given a reward function and human input, the RL algorithm may initially learn a policy that conforms to the human's instructions, but eventually might learn a policy that solely maximizes cumulative reward. This may satisfy a goal of decreasing training time since the human will have shown the RL agent high-earning states earlier than it would have explored, but the policy may still be baffling to a non-expert. The human teacher may feel like their instructions were disregarded in the long-term, creating feelings of frustration and powerlessness when they cannot directly control the agent's policies. 


Newtonian Action Advice can be seen as a way to combine supervised and reinforcement learning. A human provides information that is used to create policies that are static to the agent (supervised learning), but can be overwritten by the human. The agent uses reinforcement learning to determine the best course of action for parts of the state space in which human advice is not given. Not only does this enable the agent to use human input to decrease training time, it also empowers the human teacher to customize how the agent performs the task. It is possible the learned policy will be near-optimal instead of optimal from an objective analysis of the cumulative reward; but the agent's performance will be in greater accordance with the human teacher's instructions, which will increase the human's satisfaction with the agent's performance.



\subsection{Choosing the friction parameter}

When calculating the algorithm's `friction' parameter, it is more straightforward to think of the parameter as the number of steps each piece of advice should persist for, $S_{des}$. Increasing $S_{des}$ causes each piece of advice to be followed for a longer time, which causes a metaphorically lower friction in the model. 

Let:
\begin{quote}
\begin{description}
\item[$S_{des}$] $(steps)$ is the desired number of steps advice should persist for
\item[$S_{min}$] $(steps)$ is the minimum number of steps advice should persist for 
\item[$S_{max}$] $(steps)$ is the maximum number of steps advice should persist for 
\item[$U$] $(steps/second)$ is the domain update rate
\item[$\Delta t_{des}$] $(second)$ is the desired time between given advice
\item[$\Delta t_{min}$] $(second)$ is the minimum time between given advice
\item[$\Delta t_{max}$] $(second)$ is the maximum time between given advice
\end{description}
\end{quote}

Equations \ref{eq:1}-\ref{eq:3} show how to calculate the minimum, maximum, and desired values of the friction parameter. Two items should be noted for $S_{min}$. First, the value of $\Delta t_{min}$ has a lower bound based on human limitations. You cannot expect people to provide advice infinitely quickly. It is nonsensical to provide advice that only lasts for a fraction of a second; if $\Delta t_{min}$ is too small, it may occur that the agent follows the advice for such a short amount of time that it is not perceivable by the human. We suggest $\Delta t_{min} \geq 0.5 (seconds)$. Second, the advice must last for at least one time step, so $S_{min} \geq 1 (step)$.

\setlength{\belowdisplayskip}{0pt} \setlength{\belowdisplayshortskip}{0pt}
\setlength{\abovedisplayskip}{0pt} \setlength{\abovedisplayshortskip}{0pt}

\begin{equation} \label{eq:1}
S_{des} = \Delta t_{des} * U
\end{equation}
\begin{equation} \label{eq:2}
S_{min} = \Delta t_{min} * U
\end{equation}
\begin{equation} \label{eq:3}
S_{max} = \Delta t_{max} * U
\end{equation}

Depending on the domain, task, and nature of the actions, we suggest a starting value of  $\Delta t_{des}$ to be between 2-8 seconds.

Equation \ref{eq:4} shows the bounds on the friction parameter. The value of $S_{min}$ has the potential to run into a hard boundary based on human limitations, while $S_{max}$ is a flexible boundary based on desired behavior.

\begin{equation} \label{eq:4}
(1 step) \leq S_{min} \leq S_{des} \leq S_{max}
\end{equation}

Instead of calculating the friction parameter using the desired time between when advice is given, the average duration of each action can be used. 

Let:
\begin{quote}
\begin{description}
\item[$\Delta a$] $(actions)$ is the desired number of actions between when advice is given
\item[$spa_{avg}$] $(steps/action)$ is the average number of steps it takes to complete an action
\item[$tpa_{avg}$] $(seconds/action)$ is the average time it takes to complete an action
\end{description}
\end{quote}

If the human teacher is instructing the agent to take primitive actions, then $spa_{avg} = 1 (step)$. If the human is providing instructions for higher order actions, then $spa_{avg} \geq 1 (step)$. An action must last for the duration of at least one time step, so $\Delta a \geq 1$. Equations \ref{eq:5} and \ref{eq:6} show expressions for $S_{des}$ depending on whether the researcher has easier access to $spa_{avg}$ or $tpa_{avg}$.

\setlength{\belowdisplayskip}{0pt} \setlength{\belowdisplayshortskip}{0pt}
\setlength{\abovedisplayskip}{0pt} \setlength{\abovedisplayshortskip}{0pt}

\begin{equation} \label{eq:5}
S_{des} = \Delta a * spa_{avg}
\end{equation}
\begin{equation} \label{eq:6}
S_{des} = \Delta a * tpa_{avg} * U
\end{equation}

If primitive actions are being used, it is likely that the ideal $S_{des}$ parameter will be fairly large because the human teacher will want a given piece of advice to be followed for several consecutive time steps. If higher order actions are being used, a smaller $S_{des}$ may be beneficial because it will already take the agent several time steps to carry out the higher order action.

\section{Method}

We validated our NAA algorithm first with oracles to test the theoretical performance of the algorithm, and then with a human-subject experiment to compare the human teacher's experience with another IML interaction algorithm, Policy Shaping. 

Many of the existing machine learning algorithms that learn from human feedback are evaluated using oracles and focus on how quickly the agent learns. While this is valuable information, it ignores other important aspects of the human-agent interaction such as how humans react to the agent. For example, an oracle will never get frustrated with the agent or confused by its actions. If the interaction method affects how frustrated people are, regardless of the underlying machine learning algorithm, then perhaps that interaction method should be avoided. To that end, interactive machine learning agents should not solely be designed to optimize theoretical learning curves from simulations, but also to create a positive experience for the human teacher. 

\begin{figure}
\centering
\includegraphics[width=0.45\textwidth]{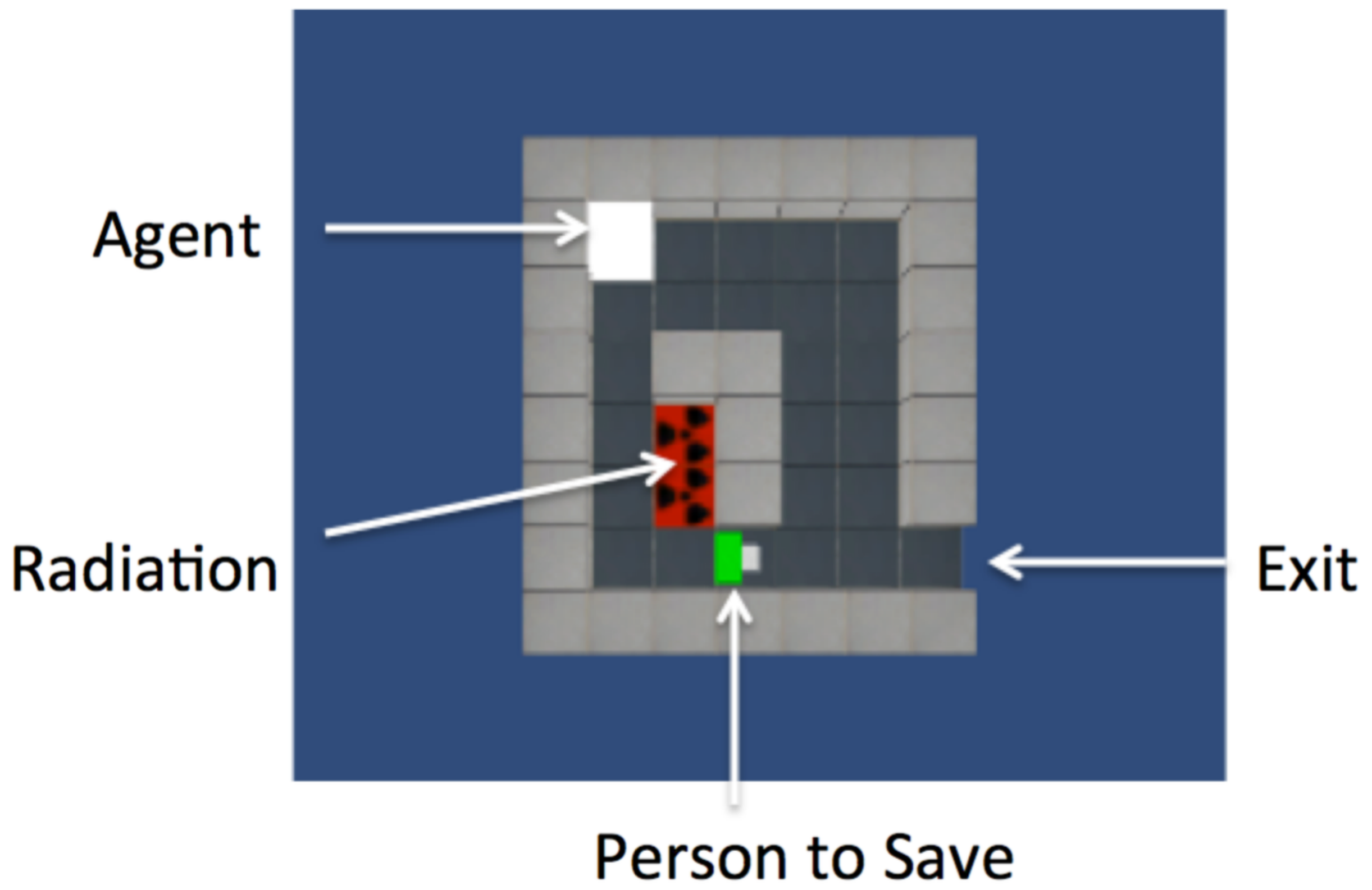}
\caption{\label{fig:radworld}Radiation World Initial Condition}
\end{figure}

Both the simulations and human-subject experiment used the same task domain. The oracle or human participant was required to teach agents to rescue a person in Radiation World, a game developed in the unity minecraft environment (Figure \ref{fig:radworld}). In the experimental scenario, there has been a radiation leak and a person is injured and immobile. The agent must find the person and take him to the exit while avoiding the radiation.

\subsection{Constructing Oracles}

We first tested the Newtonian Action Advice algorithm with simulations that used a constructed oracle to simulate human feedback. Each oracle was instantiated with a probability, $p_{advice}$, that determined how often to check for advice from the oracle. If the simulation was testing the algorithm's performance with $p_{advice} = 20\%$, then at every time step a random decimal, $d$, would be chosen between 0 and 1. The agent would check for advice if $p_{advice} \leq d$, and would otherwise not check for advice. We provided the advice for the oracles to test several cases, including maximum friction, two cases of minimal advice, and decreased friction. 

The same oracle algorithm controlling when to check for instructions was used to test the Policy Shaping agent. The same advice dictionary was used for the NAA and Policy Shaping oracles. The advice dictionary was converted to critique for the Policy Shaping agent in the following manner: if the agent took the advised action for the state, the critique was positive; otherwise, the critique was negative.

\subsection{Human-Subject Experiment}
We conducted a repeated measures human-subject experiment in which we investigated the effect of two different interaction methods, NAA and Policy Shaping, on the human's experience of teaching the agent. Both the interaction methods shared the same underlying Bayesian Q-learning algorithm.



Each participant trained two agents with different interaction methods: NAA and Policy Shaping. Both agents learned from verbal natural language instruction, which was transcribed to text using ASR software. After language processing, the processed human instructions were sent to an interaction algorithm (Policy Shaping or NAA). Then, the interaction algorithm worked with the RL agent to determine action selection. 

The Policy Shaping agent learned by incorporating a human teacher's positive and negative critique. People were instructed to provide positive or negative critique in response to the agent's actions. We used sentiment analysis as a filter to enable people to provide verbal critique without restricting their vocabulary. For example, a participant could give varied critique such as, ``Good job'', ``That's great'', ``That is a bad idea'', and ``You're wasting time''. 

The NAA agent learned from a human teacher's action advice. Participants were instructed to tell the agent to move in a desired direction. For example, if participants wanted the agent to move right, they should say, ``right.'' The only four words the participants should have used while training the action advice agent were, ``up,'' ``down,'' ``left,'' and ``right.'' These four directions were grounded to the agent's actions. 

The experiment collected data from 24 participants. None of our participants were ML experts. We made a concerted effort to recruit non-technical participants. Our participants included a piano teacher, lawyer, director of photography, political science student, and Marine veteran. 

The experiment randomly split participants into two groups. The first group trained the Policy Shaping agent first, and the second group trained the NAA agent first. Participants were told to stop training when either the agent was performing as they intended or the participant wanted to stop for any reason. Thus, the training time varied for each participant and interaction method. After participants finished training an agent, they filled out a questionnaire concerning the experience. After training both agents, the participants filled out a questionnaire directly comparing the two agents. 

In these questionnaires, the participants scored frustration, perceived performance, transparency, immediacy, and perceived intelligence. For example, immediately after training an agent, the participants were asked to score the intelligence of the agent on a continuous scale from [0:10]. A value of 0 indicated that the agent was not intelligent, while 10 meant very intelligent. The same scale of [0:10] was used for additional human factors metrics including performance, frustration, transparency, and immediacy. Values of 0 corresponded to poor performance, low frustration, non-transparent use of feedback, and a slower response time. Values of 10 meant excellent performance, high frustration, clear use of feedback, and an immediate response time, respectively. 

\section{Results and Discussion}

\subsection{Simulations}

\subsubsection{No generalization through time (extreme friction)} \label{subsec:extremeFric}

\begin{figure}[h]
  \centering
  \includegraphics[width=0.25\textwidth]{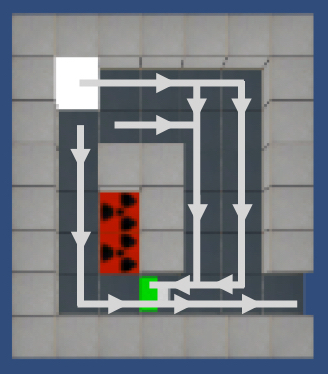}
  \caption{\label{fig:advice_maxFric}Advice given to simulation to avoid radiation.}
\end{figure}

We simulated how the percentage of time advice is followed impacts performance. The simulation was set up so the NAA agent did not generalize a given piece of advice to other states immediately after the advice was given, meaning that one piece of advice counted for only one time step (maximum friction with $S = 1 (step)$). The oracle was built with advice given for every square in the grid (Figure \ref{fig:advice_maxFric}).

\begin{figure*}[ht]
  \centering
  \subcaptionbox{Reward}[.45\linewidth][c]{%
    \includegraphics[width=.4\linewidth]{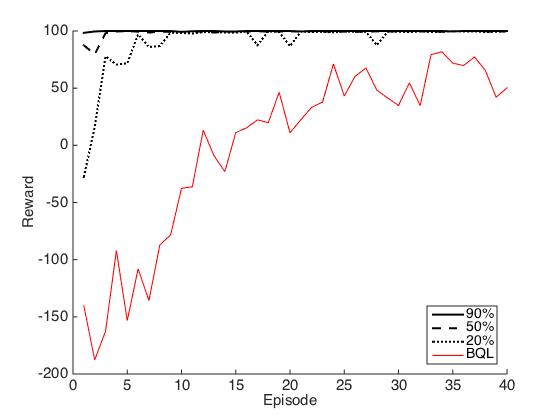}}\quad
  \subcaptionbox{Number of Steps to complete an episode}[.45\linewidth][c]{%
    \includegraphics[width=.4\linewidth]{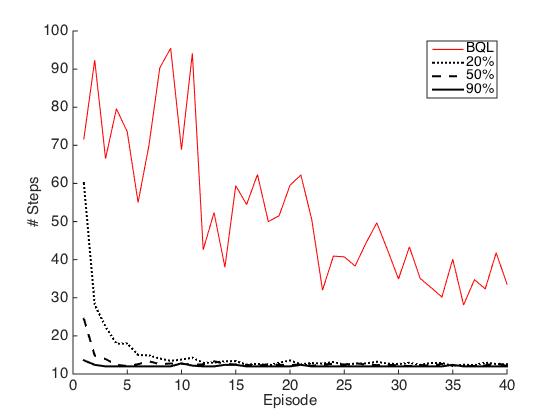}}\quad

	\caption{\label{fig:percentAdvice}How the amount of advice provided impacts performance. (advice given 20, 50, and 90 percent of the time)}
\end{figure*}

Incorporating human instruction by using the Newtonian Action Advice algorithm allows the RL agent to achieve a higher level of performance in many fewer episodes than without human input. As advice is given for a greater number of individual times steps (increasing from 20\% to 90\%), the agent accumulates more reward and completes each episode with fewer actions (Figure \ref{fig:percentAdvice}). The case with no human input is shown as BQL on the figures.

\begin{figure*}[ht]
  \centering
  \subcaptionbox{\label{fig:minAdv_minSteps}Minimal advice given to simulation to complete task with minimal steps.}[.45\linewidth][c]{%
    \includegraphics[width=.25\linewidth]{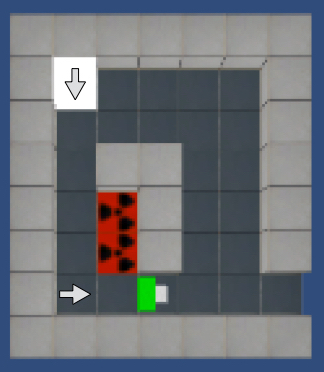}}\quad
  \subcaptionbox{\label{fig:minAdv_avoidRad}Minimal advice given to simulation to avoid radiation.}[.45\linewidth][c]{%
    \includegraphics[width=.25\linewidth]{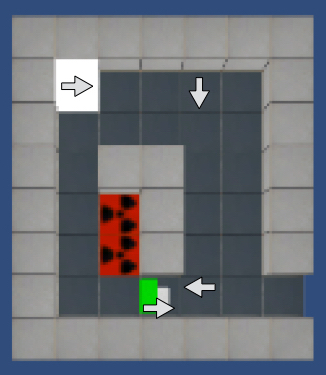}}\quad

	\caption{Minimal advice for two paths.}
\end{figure*}

\subsubsection{Minimal Advice - shortest path} \label{subsec:minAdv_minPath}

The minimal advice to take the shortest path (which takes the agent next to the radiation) is comprised of only two pieces of action advice equivalent to a human saying, ``First move \textit{down}. Then go \textit{right}.'' The minimal advice used to create the oracle in this case is represented in Figure \ref{fig:minAdv_minSteps}.

Given only two pieces of advice, the NAA agent was able to complete the episode in 10 steps achieving a reward of 102.0 every single episode. The NAA agent was set to follow each piece of advice for $S = 5 (steps)$ before returning to the BQ-L baseline action selection. 

The NAA model with a decreased friction parameter is what allows a human teacher to say ``down, right,'' instead of, ``down, down, down, down, down, right, right, right, right, right.'' It makes for a much better and more intuitive user experience to provide less instruction and not have to constantly repeat advice.

\subsubsection{Minimal Advice - avoiding radiation} \label{subsec:minAdv_avoidRad}

The minimal advice to take a path that avoids the radiation is comprised of only four pieces of action advice equivalent to a human saying, ``First move \textit{right}. Then go \textit{down}. Move \textit{left} then immediately \textit{right} after rescuing the injured person.'' This advice, which was used to construct the oracle for this case, can be seen in Figure \ref{fig:minAdv_avoidRad}.

Given only four pieces of advice, the NAA agent was able to complete the episode in 12 steps achieving a reward of 100.0 every single episode. The NAA agent was set to follow each piece of advice for $S = 5 (steps)$ before returning to the BQ-L baseline action selection. 

This case is an example of the optimal vs. customized discussion in Section \ref{subsec:customize}. The learned policy was near-optimal instead of optimal from an objective analysis of the cumulative reward since the path to avoid the radiation was slightly longer, but the agent's performance was in accordance with the human teacher's advised path.

\subsubsection{Generalization through time (friction effect)} \label{subsec:fricEffect}

We studied how the algorithm performs as the friction of the NAA model is decreased (i.e. the $S$ parameter is increased). Sections \ref{subsec:minAdv_minPath} and \ref{subsec:minAdv_avoidRad} have already shown that a small amount of advice paired with a lowered friction can enable the NAA agent to perform optimally or near-optimally from the very first episode. To test the friction effect more rigorously, we built an oracle with the same advice given for every square as Section \ref{subsec:extremeFric} (Figure \ref{fig:advice_maxFric}).

When advice is given 20\% of the time, the agent with a lower friction $S = 5 (steps)$ initially performs better than a higher friction $S = 1 (step)$. However, in later episodes the agent with lower friction earns a lower cumulative reward while taking more steps to complete each episode compared to the high friction agent. In general, these results indicate that, while lowering the friction can increase initial performance, it can also cause a lower-performing policy to be learned by the agent. 

But what is really going on in this case? We have seen in Sections \ref{subsec:minAdv_minPath} and \ref{subsec:minAdv_avoidRad} that minimal advice paired with a lowered friction enables the agent to perform optimally or near-optimally from the very first episode. Why would providing more advice ($p_{advice}=20\%$) harm the agent's performance, particularly when Section \ref{subsec:extremeFric} showed that increasing advice increases performance? The core issue is a limitation of the oracle. At every time step, the oracle listens for advice with a probability of 20\%. This is not how a human would provide advice. The decreased performance in this case occurs when the agent spends time repeatedly banging into walls after advice has been generalized to a wall state instead of the oracle providing advice for that wall state. The probability $p_{advice}=20\%$ is low enough that this behavior is not corrected for many episodes. Human teachers who observed this behavior would quickly provide an extra piece of advice to make sure the agent did not fruitlessly waste time. 

When people decrease advice, they tend to limit themselves to the most important pieces of advice, such as the minimal advice cases. The oracle has no way to know which advice is the most important, and so provides advice in a way that is not indicative of human behavior. A possible solution to this problem is to build more elaborate oracles that more accurately represent human behavior. There are three main issues with this approach: 1) the use of and response to an algorithm will vary across individuals, so multiple contradictory oracles would need to be constructed, 2) an oracle's ability to provide a type of input does not mean a human is likely or able to provide that input in reality, and 3) it is very unlikely that even the most elaborate oracle could simulate the human's response to the agent, such as frustration. A more practical solution to this problem is to test algorithms with human-subject experiments.

This case shows why IML researchers should verify interaction algorithms with human-subject experiments in addition to simulations. If we had analyzed these results without understanding the limitations of the oracle, we might have discarded parameterizations using a lower friction.

\begin{figure*}[ht]
  \centering
  \subcaptionbox{\label{fig:compFric_rew}Reward}[.45\linewidth][c]{%
    \includegraphics[width=.4\linewidth]{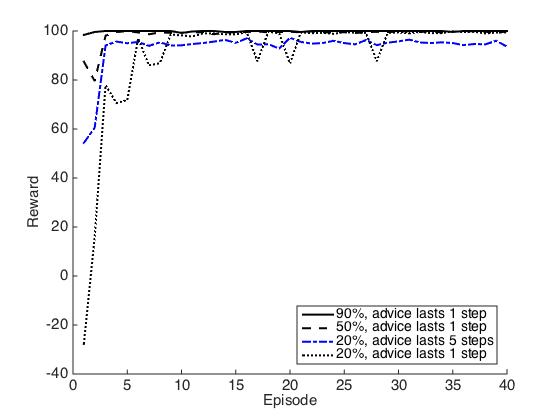}}\quad
  \subcaptionbox{Number of Steps to complete an episode}[.45\linewidth][c]{%
    \includegraphics[width=.4\linewidth]{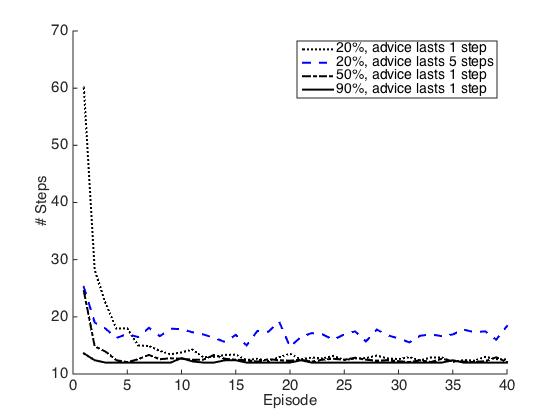}}\quad

	\caption{Generalization through time. (advice given 20, 50, 90 percent of the time)}
\end{figure*}



\subsubsection{Comparison of Newtonian Action Advice and Policy Shaping}

\begin{figure*}[ht]
  \centering
  \subcaptionbox{\label{fig:naaVsPS_rew}Reward}[.45\linewidth][c]{%
    \includegraphics[width=.4\linewidth]{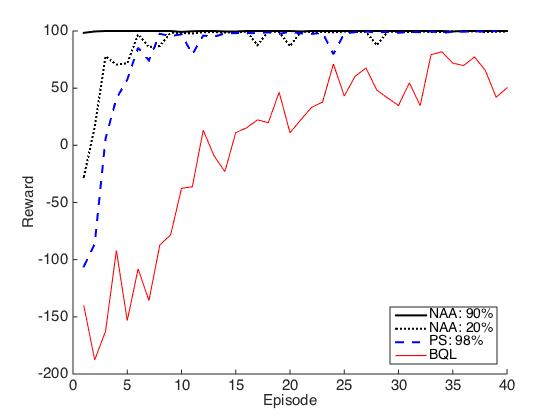}}\quad
  \subcaptionbox{Number of Steps to complete an episode}[.45\linewidth][c]{%
    \includegraphics[width=.4\linewidth]{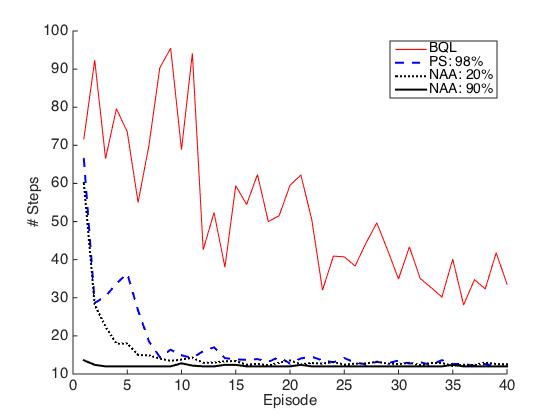}}\quad

	\caption{\label{fig:oracle_naaVsPS}Comparison of Newtonian Action Advice and Policy Shaping}
\end{figure*}

Figure \ref{fig:oracle_naaVsPS} shows that, given equivalent input, Newtonian Action Advice can learn faster using less human instruction than Policy Shaping. Even when Policy Shaping used advice 98\% of the time, it learned slower than the NAA agent that was using input only 20\% of the time. The oracle used the same setup as Sections \ref{subsec:extremeFric} and \ref{subsec:fricEffect} (Figure \ref{fig:advice_maxFric}).


When learning from human teachers in practice, however, the performance of each agent is entirely dependent on the instruction provided by the person. Neither agent is guaranteed to perform better than the other. If the human provides no instructions, the Policy Shaping and NAA agents perform equally since they both reduce to a Bayesian Q-learning algorithm. In order to investigate how the performance of the agents varied with real human teachers, as well as how the human experience was impacted by interacting with each agent, we conducted a human-subject experiment.

\subsection{Human Subject Results}

\begin{figure}[h]
  \centering
  \includegraphics[width=0.8\textwidth]{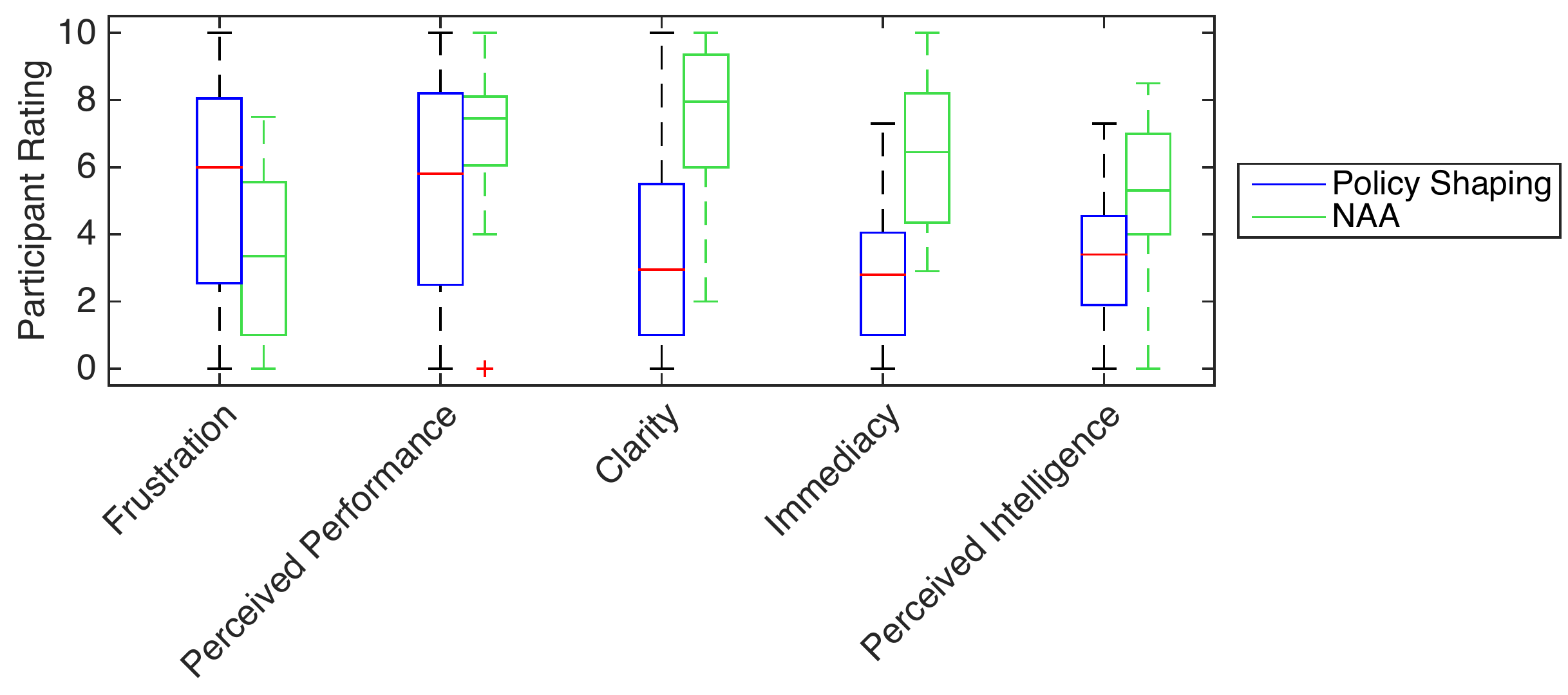}
  \caption{\label{fig:hf_comp}Comparison of the human experience.}
\end{figure}

Immediately after training each agent, participants were asked to score aspects of their experience training the agent, including frustration, perceived performance, transparency, immediacy, and perceived intelligence (Figure \ref{fig:hf_comp}). Paired t-tests were conducted for each metric in which the null hypothesis was the pairwise difference between the two paired groups had a mean equal to zero. We found that all measured aspects of the human experience differed significantly between the two agents. 

In summary, compared to Policy Shaping participants found the Newtonian Action Advice agent to be less frustrating, clear and more immediate in terms of how the agent used human input, better able to complete the task as the person intended, and more intelligent. 

In addition to creating a better human experience, the NAA agent also performed better than Policy Shaping in terms of objective RL metrics. The average training time was smaller for the NAA agent. The number of steps the agent took to complete each episode was smaller for NAA. The average reward was higher for NAA than Policy Shaping. However, the amount of input provided by the human teachers was statistically equal for the two interaction algorithms.

\section{Conclusions}

This paper presented Newtonian Action Advice, a method to integrate a human's interactive action advice (ex: ``move left'') with reinforcement learning. For equivalent human input, Newtonian Action Advice performs better than Policy Shaping, both in terms of RL metrics like cumulative reward and human factors metrics like frustration.

\newpage

\subsubsection*{Acknowledgments}

This work was funded under ONR grant number N000141410003.


\bibliographystyle{apalike}
\bibliography{tmpSample}

\end{document}